# Internet of Things Networks: Enabling Simultaneous Wireless Information and Power Transfer

Prerna Dhull[1,2], Andrea P. Guevara[2], Maral Ansari[3], Sofie Pollin[2], Negin Shariati[1], and Dominique Schreurs[2]

[1]RF and Communication Technologies (RFCT) Lab, University of Technology Sydney, Australia

[2]Div. ESAT-WAVECORE, KU Leuven, Belgium

[3]Global Big Data Technologies Center (GBDTC), University of Technology Sydney, Australia

## 1  Introduction

The number of sensors deployed in the world is expected to explode in the near future. At this moment, nearly 30 billion Internet of Things (IoT) devices are connected and this number is expected to double in the next four years. While not all of these are battery powered, as technology becomes smaller and mobility becomes more important to consumers, soon a larger portion will be. This forecast predicts that the number of machine to machine (M2M) devices will have the largest increase, representing nearly 50% of all devices in 2023 [1, 2]. These devices are typically small and therefore ideal candidates to be powered wirelessly. Typical examples are healthcare monitoring, smart homes, and industrial applications.

As IoT devices are embedded in Wireless Sensor Networks (WSNs) and communicate with a Base Station (BS), the logical step is to have the BS transmit both power and information wirelessly to the IoT devices using the same Radio Frequency (RF) signal, as demonstrated in Fig. 1, while the sensor node performs information detection and energy harvesting operations over the same RF signal. This approach has been termed simultaneous wireless information and power transfer (SWIPT), from Wireless Information Transfer (WIT) and Wireless Power Transfer (WPT). In WIT, a signal is utilized only for information transmission, whereas in WPT, the signal is used only to deliver power to the node. SWIPT provides a bridge between these two technologies by exploiting the same signal for providing both information and power to IoT devices. In this article, we focus on downlink WIT only. The IoT sensor node would typically send sensed data to the BS, using backscattering or another approach, representing uplink WIT. Further, as seen in Fig. 1, there has to be a power management unit between the information receiver and power storage unit, which consumes very low power levels [3]. However, in this article, this is considered as part of the power storage system in order to keep the receiver architecture representation simple.

From a WPT perspective, it would be beneficial to transmit the RF signal only to the IoT nodes which need power, avoiding wasting power on the rest of the IoT nodes. The solution to this can be beamforming. Beamforming, or spatial filtering, is a technique used in arrays for directional signal transmission (or reception). Beamforming achieves this property by coherently combining the fields radiated by the array elements, to direct their radiated energy into par-ticular directions. These multiple beams are created at the BS to communicate with different IoT nodes simultaneously [4].

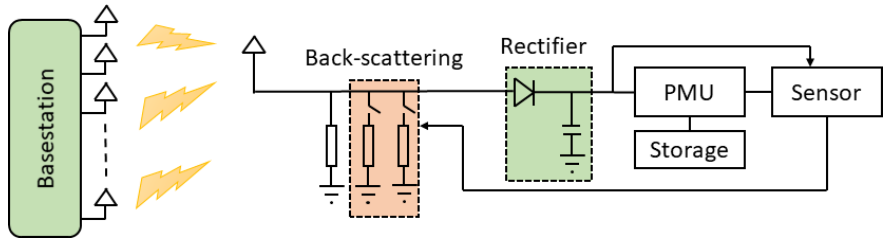

Fig. 1: Base station transmitting to a sensor node consisting of rectifier, PMU (power management unit), storage unit for power extraction and backscattering for information transfer from the received signal.

In a wireless communication system, antenna characteristics that have high directivity and wide beamwidth can be used for long distance and multi-user coverage. BS antennas for WPT with such performance can broadcast and deliver wireless power to a large number of widely distributed IoT users. However, omni-directional antenna transmission, which radiates in all directions, increases the interference. In such situations, the use of narrow directional beams enhances the signal to interference noise ratio by sending the signal toward the specific user, and achieves higher data rates by reducing multipath effects and interference [5–7], as seen in Fig. 2. In Table I, some of the pros and cons of using beamforming are summarized. On balance, beamforming not only enhances WIT performance but also WPT performance, and this makes it a suitable candidate for SWIPT systems.

Compared to conventional transmission design, implementation of beamforming techniques requires additional signal processing at the transmitter. Further, beamforming utilizes large antenna arrays for steering the narrow beams to enlarge the coverage area, which results in high cost and high-power consumption systems. Besides, employing such large arrays makes the systems bulky and less efficient than traditional systems. However, with the recent advances in hybrid antenna arrays, employing a combination of analog and digital beamforming techniques serves as a cost and energy efficient alternative. Therefore, today's beamforming can serve as a potential solution for SWIPT where it is expected to improve WPT as well as WIT.

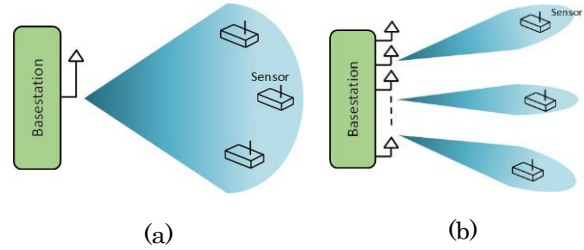

Fig. 2: Illustration of a BS coverage using (a) single antenna compared to adopting (b) a beamforming antenna array.

Table I: Beamforming deployment in wireless networks.

| Advantages | Disadvantages |
| --- | --- |
| • Increased capacity | • Higher complexity |
| • Reduced interference | • Higher cost |
| • Increased coverage area | • Larger sizes |
| • Improved transmission efficiency | |

In addition to signal covering capabilities, it is essential to take into account the IoT node architecture. Prior to SWIPT, receiver structures were being studied separately either for wireless communication purposes only or for energy harvesting purposes only, whereas now these operations need to be performed on the same signal at the IoT node side. Therefore, the problem of integrating these two processes in an efficient manner arises [8]. Initially, in 2013, two types of receiver setups were introduced as shown in Fig. 3: (a) a separated information-energy receiver architecture where the signal is divided between two streams, one for information decoding and the other for energy harvesting. Various signal splitting techniques can be considered, such as time switching (TS) and power splitting (PS); (b) an integrated information-energy receiver where signal splitting is performed only after passing the whole signal through the rectifying circuit (i.e., no RF mixer is required at the receiver [9]).

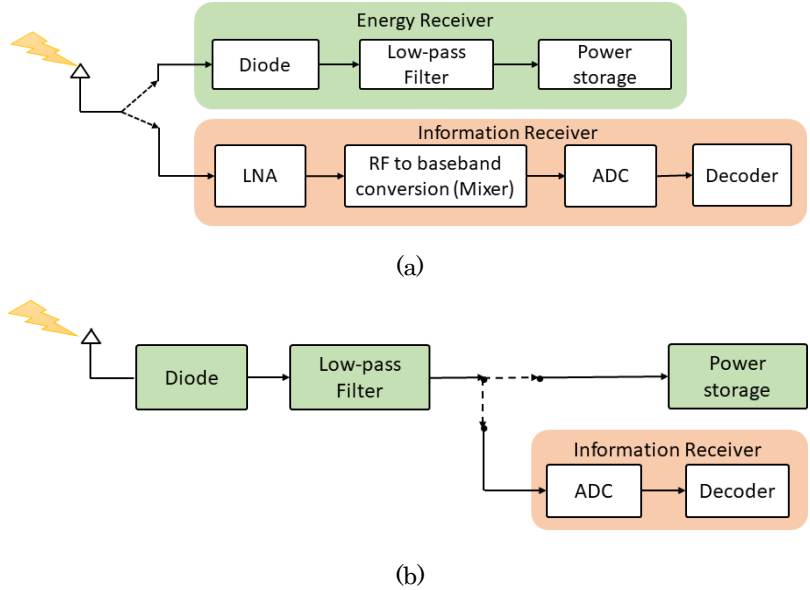

Fig. 3: SWIPT architectures: (a) Separated information-energy receiver and (b) Integrated information-energy receiver [9].

For both SWIPT architectures, there is a trade-off between the achievable information rate and the amount of harvested power, as both metrics cannot be maximized simultaneously [9]. Further, the use of an energy harvester (rectifying circuitry consisting of diodes) at the receiver introduces non-linearity into the SWIPT system, and this non-linearity significantly affects the power conversion efficiency (PCE) performance at the output [10]. Due to this non-linearity, the output power is not only a function of received signal power, but also of the received signal shape. For example, a waveform having higher peak-to-average power ratio (PAPR) with the same average input power is able to turn-on the diode earlier, resulting in an increased PCE at the output.

However, these high PAPR waveforms deteriorate the WIT performance because of saturation of the non-linear amplifier at the transmitter. From a SWIPT system perspective, the performance of both the WIT and the WPT is important and it is therefore necessary to maximize the rate-energy trade-off at the receiver. Thus, for a complete SWIPT system, we need to co-design not only the transmitter and receiver hardware, but also the transmitted waveform for a particular type of IoT node architecture in order to enhance the SWIPT performance.

Based upon the above discussion, this article reviews two questions related to the BS: (1) how can a BS utilize beamforming for sending power to specific IoT devices while avoiding wasting power in zones where there are no IoT devices, and (2) how can signal transmission by a BS be optimized for a particular IoT receiver architecture so that the strongly energy-constrained IoT node can receive both information and power efficiently?

## 2   Beamforming at Base Station

A BS can achieve beamforming in multiple ways. Digital beamforming, including multiple input-multiple output (MIMO) signal processing, is the most flexible approach to control the power direction. However, conventional fully digital beamforming with one dedicated RF chain for the transmit antenna is too costly in terms of power consumption, as it is necessary to have a separate power amplifier for each RF chain. In addition, all the antennas must be perfectly synchronized to avoid signal distortion during uplink and downlink data transmission. A cost and energy efficient solution is analog multiple beam antenna arrays. Most such antennas can be phased to synthesize a broad cov-erage beam, but the energy transmitter needs a very large number of antennas and adaptive beamforming to flexibly control the power direction. However, hybrid beamforming is able to achieve the same optimal performance as fully digital beamforming, as long as the number of RF chains at the transmitter is no less than twice the number of sub-bands used or twice the number of channel paths [11].

In the following, we will first review approaches for analog beamforming that are applicable for SWIPT. Due to the steep increase in IoT nodes, BSs are evolving toward a massive number of antennas, adopting digital beamforming or, as mentioned, a hybrid digital/analog approach. An experimental study mimicking 120 users is presented in the second subsection.

## 2.1  Analog Beamforming

There are different types of analog beamforming that can be used in wireless applications. These include circuit-type [12] and quasi-optical beamforming networks (BFNs) [13], the latter of which are more desirable for higher frequency communication systems. In this article, circuit-type BFNs are presented that have higher significance for SWIPT.

A BFN is a physical layer element of an array system that combines signals with the requisite amplitudes and phases required to produce a desired angular distribution of the emitted radiation (i.e., one or more beams pointing in prescribed directions). Energy delivered to a particular input port of the BFN is thus associated with a particular beam radiated by the antenna elements connected to the BFN's output ports [4]. In the following, the two most-used circuit-type BFNs are introduced.

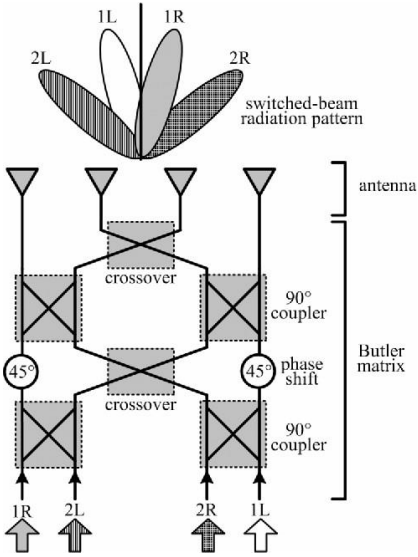

Fig. 4: Schematic representation of the conventional 4 × 4 Butler matrix [18].

### 2.1.1 Butler Matrix

The Butler matrix was first described by Jesse Butler and Ralph Lowe in [14]. The Butler matrix offers a set of orthogonal beams, with the beam directions dependent on frequency. The $M \times N$ Butler matrix consists of $M$ number of inputs that independently feed $N$ number of outputs with progressive phase delays. Inputs and outputs are well matched and isolated, and the network is theoretically lossless [15]. There is no beam spacing loss due to the nature of orthogonal beams [16,17]. The outputs of the Butler matrix can be connected to the antenna array to produce the required amplitude and phase for the antenna elements to enable beam switching capabilities.

Consider a Butler matrix as a microwave networks with 4 input ports and 4 output ports as shown in Fig. 4. The network consists of several microwave components, including couplers, phase shifters, and crossovers. The network has the special characteristic that if a signal is applied to input $i$ ($i$ = 1, 2, 3, 4) then the outputs all have equal amplitude and output $j$ ($j$ = 1, 2, 3, 4) has phase $360(j-1)(i-1)/n$ degrees, which means that feeding element $i$ radiates a beam at $sin^{-1}\frac{\lambda i - 1}{sn}$ azimuth, where $s$ is the spacing of the columns. To produce more focused and narrower beams as required for SWIPT, high-gain antenna arrays are necessary. Array gain is the increase in the average signal-to-noise-ratio (SNR) obtained by combining multiple antenna elements compared to a single element. In order to increase the gain of the array, the Butler matrix outputs $i$ must increase. The amplitude of these outputs should have unequal powers to achieve lower sidelobes because of nonuniform illumination. A prototype Butler matrix design covering the above requirements is reported in [19]. The reported multi-beam antenna can create four beams in the azimuth plane. The measured radiation patterns show that the beams cover a spatial range of roughly 90°. The design architecture and the final prototype are shown in Fig. 5. Recently, a compact $4 \times 4$ Butler matrix for IoT applications was presented in [20]. The structure operates from 2.35 to 2.55 GHz, which covers the industrial, scientific, and medical bandwidth (BW). Small dimensions of $31.3 \times 22.9$ mm make this design useful for IoT applications.

### 2.1.2 Nolen Matrix

The Nolen matrix was first proposed by John C. Nolen [21]. In Fig. 6, the schematic diagram of an $M$-entry Nolen matrix is shown. A Nolen matrix is a lossless variant of a Blass matrix in which all the directional couplers below the diagonal have been removed. Consequently, all the matched loads at the end of the feeder lines have also been removed. In a Blass matrix any signal present above the diagonal will necessarily be dissipated in the matched loads. To eliminate these losses, no signal should go beyond the diagonal of the directional coupler matrix. This can be obtained along the diagonal with directional couplers with coupling values of 0 dB, which is the functional schematic of the Nolen matrix. In summary, the Nolen matrix may be considered as an asymp-totic singular case of a Blass matrix for which the design constraint is on the values of directional couplers. Nolen matrices can be used for WPT beamforming applications. A symmetrical 3 × 3 uniplanar Nolen matrix is introduced in [22], where the design reduces the number of phase shifters and does not require outside phase compensation lines. The design architecture and the final prototype of the 3 × 3 Nolen matrix feeding network are shown in Fig. 7.

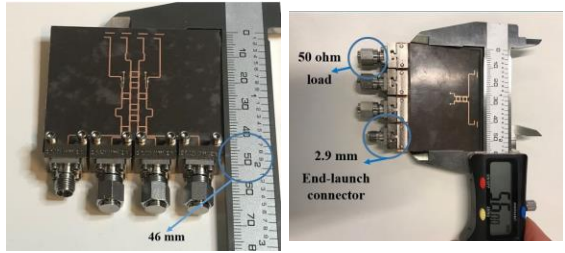

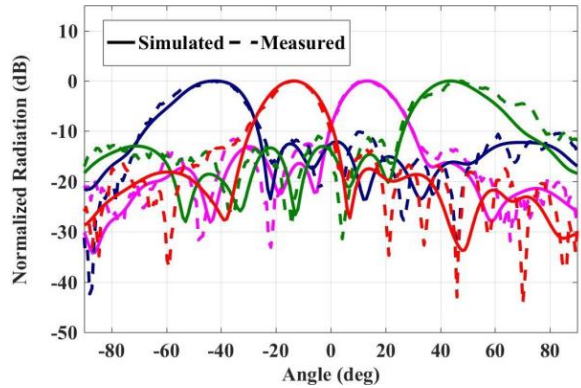

Fig. 5: Photographs of the fabricated four-beam antenna array using conventional 4 × 6 Butler matrix [19]. (a) Top view. (b) Bottom view. (c) Simulated and measured beams.

In conclusion, employment of analog beamformers is an energy efficient and low-cost solution for realizing a high directivity linear array with narrow beamwidth in a one principal plane. These are one form of cost-effective implementation of beamforming on a PCB substrate, while benefiting from compact size. In [24], the authors introduced practical application examples of such beamformers for broadcasting wireless power to multiple IoT devices that are remotely located. The array broadcasts wireless power to multiple IoT targets at different elevations.

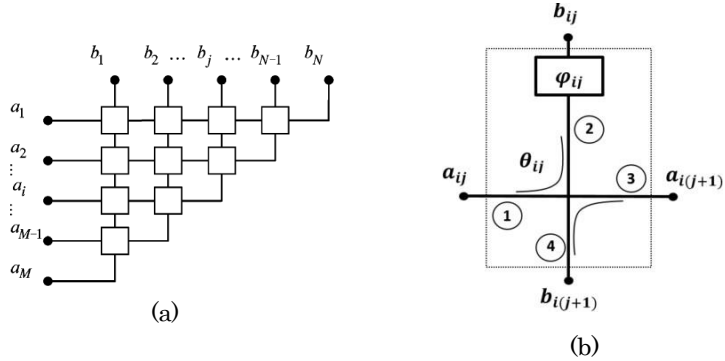

(a)

(b)

Fig. 6: (a) Functional schematic representation of *M*-entry Nolen matrices (d) Details of each node [23].

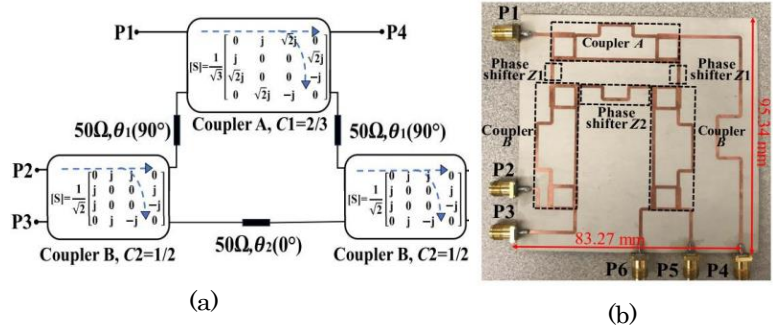

(a)

(b)

Fig. 7: 3 × 3 Nolen matrix with lumped-element couplers and phase shifters (a) Simulated layout, and (b) Fabricated photograph [22].

Analog beamformers can be used to enhance the end-to-end power transfer efficiency from the transmitter to the receiver. Conventional systems achieved this by using bulky structures such as parabolic reflectors where the antenna needs to be mechanically adjusted to focus the signal toward the receiver.

## 2.2 Digital Beamforming

In wireless IoT networks, BSs have to communicate with and send power to a high number of nodes. For this reason, BSs are evolving toward massive MIMO architectures, meaning that a large number of antennas are deployed. Those antennas can be allocated in a centralized or distributed manner [25, 26]. In communications, the power leakage in space can lead to interference when users are active in those areas. Therefore, different digital beamforming techniques such as maximum-ratio transmission/maximum-ratio combining (MRT/MRC) [27], zero-forcing (ZF) [28], regularized zero-forcing (RZF) [29], and minimum mean squared Error (MMSE) [30] are used to nullify unwanted areas. The differences among these beamforming methods are summarized in Table II.

Few experimental studies analyze the impact of power leakage in massive MIMO systems. In [31], two off-the-shelf receivers measure the leakage power in a centralized massive MIMO system. The difference between them is whether the wireless channel knowledge is present at the receiver or not. This work presents the measured harvested output voltage as a function of the linear distance (up to 1 meter) between the target user and the leakage receiver. Although the number of samples is limited, this work provides initial experimental insight into the potential use of massive MIMO and digital beamforming for power transfer.

A more extensive measurement campaign was conducted at KU Leuven [26, 32, 33]. An automated massive MIMO experiment analyzed three-antenna configurations and the corresponding leakage power impact in an indoor scenario. The power leakage is approximated based on the wireless channel knowledge collected for 120 locations.

Table II: Beamforming techniques for wireless networks.

| Precoding/ Combining Vector | Advantages | Disadvantages |
|---|---|---|
| Maximum-Ratio Transmission/ Maximum-Ratio Combining (MRT/MRC) [27] | • Lower computational complexity.<br>• Maximizes the power directed toward the desired user/sensor. | • Does not actively suppress leakage.<br>• Provides the lowest spectral efficiency and capacity for a massive MIMO system. |
| Zero-Forcing (ZF) [28] | • Actively suppresses leakage. | • Trade-off between leakage suppression and power directed toward the desired user/sensor.<br>• Does not work in systems with a lower SNR. |
| Regularized Zero-Forcing (RZF) [29] | • Actively suppresses interference plus noise. | • Requires an estimation of the system SNR. |
| Minimum Mean Squared Error (MMSE) [30] | • Provides the optimal performance of the system by suppressing intra and inter-cell interference and noise. | • Computationally costly.<br>• Requires knowledge of the channels of all the users that are served and not served by the BS. |

### 2.2.1 Experimental Setup

The KU Leuven massive MIMO testbed used for this measurement campaign comprises two elements: a 64-patch antennas BS (Fig. 8) and two Universal Software Radio Peripherals (USRPs), by which each USRP is associated with two User Equipment (UE). The modular patch antennas allow different configurations. In this case, we evaluate the following three antenna array topologies:

- Uniform Rectangular Array (URA): The 64-patch antennas are deployed in a rectangular array of $8 \times 8$. See Fig. 8(a).

- Uniform Linear Array (ULA): The 64-patch antennas are lined up next to

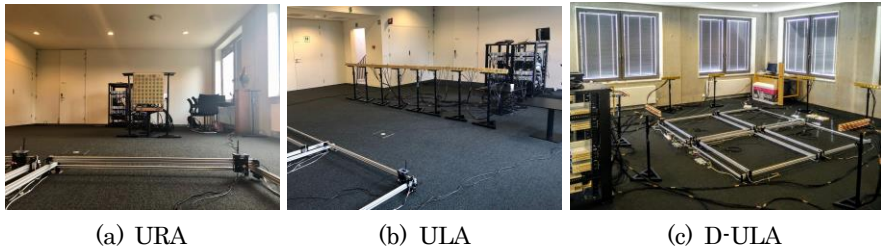

(a) URA          (b) ULA          (c) D-ULA

Fig. 8: KU Leuven massive MIMO antenna configurations.

each other, in an horizontal array of $1 \times 64$. See Fig. 8(b).

- Distributed Uniform Linear Array (D-ULA): The 64-patch antennas are distributed around the analyzed area in 8 ULA arrays of $1 \times 8$ each. See Fig. 8(c).

Each UE controls two independent uplink data streams associated with two dipole antennas. The uplink channel is estimated with these data streams. The four dipole antenna elements are connected to four positioning systems that can freely move in a grid of 1.3 m $\times$ 1.3 m each. All the positioners move synchronously 20 cm horizontally and 30 cm vertically, then stop for ten seconds. The wireless channels between the four dipole antennas (at the UEs) and the 64-patch antennas (at the BS) are collected during this period. During post-processing, the wireless channel of each position is treated as an independent virtual user. In this way, we can create a massive MIMO scenario with 120 users deployed on the ground [34, 35].

### 2.2.2 Experimental Results

Fig. 9 shows the heatmap with contours for the experimental setup for a different number of active antennas at the BS. The target user is located in the center of each figure. The digital beamforming MRT is applied toward the target user. Those figures represent the estimated leakage in the studied area as the normalized received signal over 0 dB.

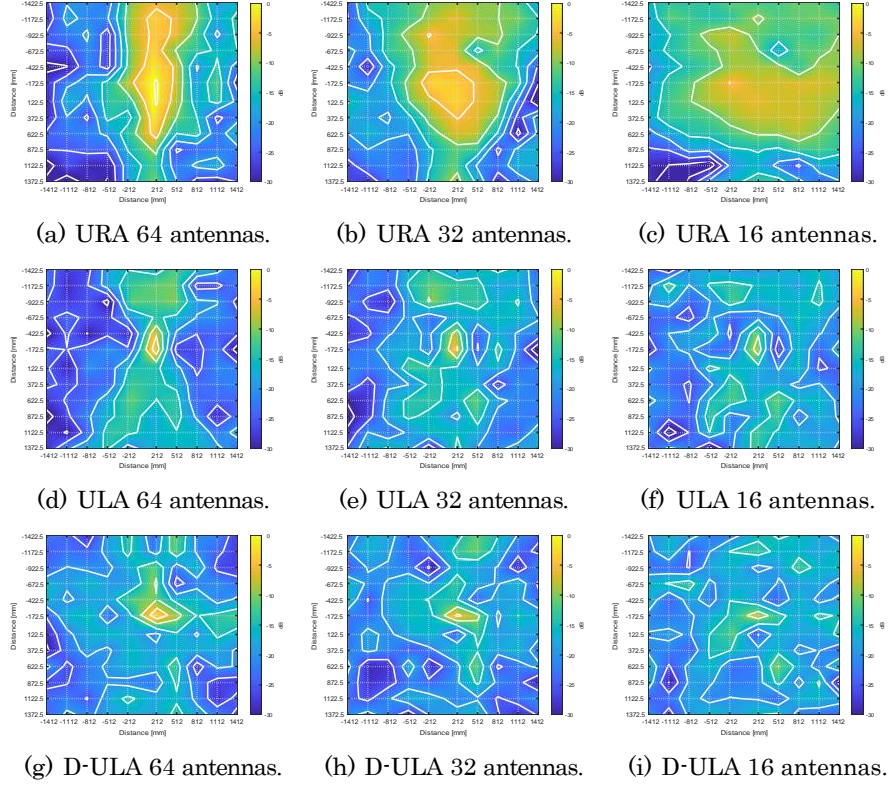

(a) URA 64 antennas. (b) URA 32 antennas. (c) URA 16 antennas.

(d) ULA 64 antennas. (e) ULA 32 antennas. (f) ULA 16 antennas.

(g) D-ULA 64 antennas. (h) D-ULA 32 antennas. (i) D-ULA 16 antennas.

Fig. 9: Contouring normalized over 0 dB which is the maximum value received by the target user in each antenna configuration regardless of the number of active antennas.

It is clear that the strongest yellow peak is visible for the target user, and the power distribution follows a beam for the URA and ULA scenarios. It is interesting to see that, with MRT beamforming, the signal amplitude of the weakest signal can be up to 30 dB lower than the peak amplitude of the strongest signal. However, these dark blue spots mainly occur in the ULA scenario. For the D-ULA topology, the spot beams are noticeable, especially for the scenario with 64 distributed antenna elements.

From the performed experiments, it can be observed that beamforming is significantly helpful in reducing the interference among users, while focusing power only upon the desired user. Therefore, beamforming can be utilized for careful power management in SWIPT systems where both information and power delivery are of critical importance.

## 3 SWIPT Signal Design at Base Station

For an overall energy efficient SWIPT system, not only is effective delivery of the RF signal power to a UE important, but it is also necessary to maximize the amount of harvested energy at the receiver, while providing a sufficient data rate. It has recently been observed that different signal waveforms result in different energy efficiencies at the receiver. Therefore, the area of SWIPT signal design with the corresponding modified receiver architectures also needs to be explored to enhance the end-to-end SWIPT performance. In this article, various SWIPT waveforms to enhance SWIPT system performance are discussed, along with an explanation of how the corresponding receiver architecture needs to be redesigned. The focus is, however, on the general receiver architecture, whereas the required SWIPT receiver implementation details can be found in [36], where both near-field and far-field SWIPT technologies are discussed.

As mentioned earlier, in SWIPT systems, it is not possible to maximize both the information transfer rate and the amount of harvested energy simultaneously, and a trade-off exists between these two objectives. For a long time, high PAPR signals have been used to enhance WPT system performance, and this has been achieved in multiple ways. One such designing a higher PAPR signal with the same average input power is to concentrate the signal power within a smaller duration instead of covering the complete duty cycle [37], and it has been shown that to obtain an output voltage of 1.6 V, only $-5.1$ dBm input power is required with a pulsed signal instead of $-3.44$ dBm for a continuous waveform.

Another way to increase the PAPR of the input signal at the receiver is to use a multitone signal instead of a single tone signal [10]. As the diode (or diodes) in the receiver node is nonlinear, the multiple tones present in the received signal undergo intermodulation, resulting in multiple frequency components in the baseband, corresponding to a time domain voltage ripple in addition to the desired DC voltage, negatively impacting the PCE performance. Therefore, the several parameters affecting this ripple, such as number of tones, tone separations, tone phases, input power range, the rectifier's low-pass filter's cut-off frequency, and diodes have to be carefully studied [38–41].

However, in a SWIPT system where both information transmission and power transfer are required, the additional ripple can actually be exploited from the information perspective. Consequently, the signal excitation design at the BS is a challenging task. In practice, the shape of the waveform transmitted by the BS may be altered before it reaches the receiver node due to multipath effects in the wireless channel. This can be mitigated with precoding and pre-equalization techniques. The former has been touched on in the previous subsection on digital beamforming. Regarding the latter, a non-dispersive wireless channel is assumed in this work for simplicity reasons.

In a communication system, the actual modulated waveform varies over time, as opposed to the deterministic multitone signal used for WPT, as streams of symbols are sent between the BS and IoT node. This randomness often results in a lower PCE compared to the deterministic multisine waveform [42].

The modulated waveform is defined by several parameters, such as the type of waveform, modulation scheme, modulation order, symbol rate, and input distribution, which can affect the energy harvesting and communication performance. The values of the parameters can be chosen to reach a trade-off, or to favor either wireless power transfer or communication, if one aspect is more critical for the practical application.

In the following, we discuss modulated waveforms for SWIPT, making a distinction between the two node receiver configurations under consideration, as presented in Fig. 3.

## 3.1 Separated Node Architecture

First we focus upon the node architecture in which the received RF signal is divided into two streams: one for the information signal and one for the power signal, corresponding to Fig. 3(a). This is usually performed with the help of a power divider with a certain power ratio for the energy harvesting performance at the receiver output. The use of a separated node architecture at the receiver offers the advantage of using existing transmission schemes such as Orthogonal Frequency Division Multiplexing (OFDM), a combination of multiple subcarriers. This requires only slight modifications in the transmitted signal and the receiver architecture. The reason for this is the use of existing communication system information detection methods, as signal splitting is performed in the RF domain for the two separate paths, and only an additional energy harvester is required at the receiver. However, this is attained at the cost of higher power consumption at the receiver due to the presence of mixers/FFT.

One way to achieve separate signals for information and energy transfer is to segregate these OFDM subcarriers into two sets, as shown in Fig. 10(a), by using well-designed band pass filters (BPF), and therefore eliminating the need for power splitting or time switching at the receiver [43]. It is possible to harvest 18 mW for 4 bps/Hz from the transmitted signal power of 1 W using this OFDM subcarrier splitting technique. However, here, the required BPFs need to be really sharp to support an OFDM signal having smaller sub-carrier

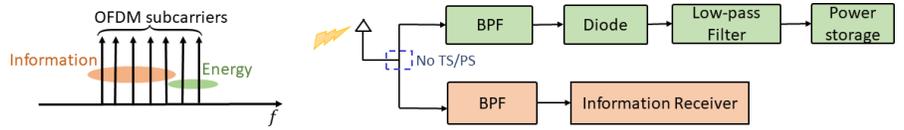

(a) OFDM frequency spectrum and corresponding receiver architecture.

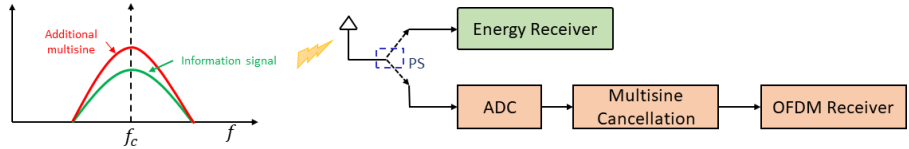

(b) Frequency spectra of multisine superimposed upon the OFDM information signal and corresponding receiver architecture.

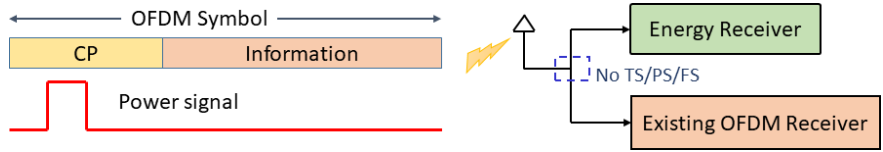

(c) CP-OFDM information symbol superimposed on the rectangular pulse power signal.

Fig. 10: Waveforms and corresponding receiver architecture for separated information-power configuration.

frequencies. This increases the computational complexity at the receiver because of the increased Discrete Fourier Transform (DFT) length [44]; consequently, increasing the power consumption at the receiver. Therefore, it is necessary to monitor the signal processing power consumption at the receiver in order to check the feasibility of this scheme.

An alternative, tailored for the PS configuration, is to superimpose an additional multisine on the OFDM waveform, as illustrated in Fig. 10(b), to further enhance its PAPR [45]. For a 20 dB SNR, information and power performance of 1 bit/s/Hz and around 1.8 µA current can be attained for the combination of OFDM and 16-tone deterministic multitone. Although the only modification needed in this receiver architecture is the incorporation of cancellation of this multisine waveform, the drawback of this approach is that the receiver's analog-to-digital-converter (ADC) may suffer from saturation when the power signal level is relatively higher than the information signal.

Another way to capitalize on the high PAPR OFDM signal for energy harvesting is to exploit its redundant cyclic prefix (CP), as shown in Fig. 10(c). In this approach, a high PAPR rectangular pulse for power transfer can be superimposed on the CP of the OFDM information signal [46]. As the cyclic prefix is discarded for information decoding, no modification is required in the information receiver, and 1.4 V can be harvested in a $-15$ dB SNR environment. However, proper functioning of this signal requires optimization of the rectangular pulse width to minimize interference with the information part of the signal. Another approach for using the redundant CP for power transfer is proposed in [47]. Here, instead of transmitting a separate power signal over the OFDM signal, some portion of the information part is utilized for energy harvesting in addition to CP and the length of this used information part for energy harvesting is optimized according to the energy requirement at the receiver.

In [48], directly superimposing a DC signal over the OFDM signal to make an OFDM-DC signal is proposed. In such a case, the power can be directly transferred in DC, in addition to the OFDM information signal. It is able to have 2 mW power and a symbol error rate (SER) of $10^{-1}$ for 10 dB SNR with 2 bits/channel of the quadrature-phase-shift-keying (QPSK) OFDM-DC signal. This technique offers the advantage of elimination of non-linear rectifying components at the receiver, which saturate the PCE in rectifier-based schemes, and a low-pass filter is sufficient for power conversion. However, the drawback of this transmission technique is that it requires a complete redesign of the transmitter and receiver architectures.

Instead of dividing the received signal stream into two signal streams using

PS or TS, dividing the symbol constellation points to enhance the SER performance of the system is introduced in [49]. Here, hybrid constellation shaping is utilized to map constellation points such that symbols closer to the origin would be recognized as information symbols and symbols farther from the origin, having larger amounts of energy, would be considered as energy symbols. It is assumed that the receiver has prior knowledge of the constellation points' functions. This approach has been shown to increase SER performance compared to the earlier PS based scheme. Another symbol designing scheme has been introduced in [50]. Here, an asymmetric QPSK with constellation points more inclined toward the zero phase value are shown to have an improved rate-energy trade-off over the general QPSK performance.

The performance analysis of the above-discussed transmission designs is summarized and compared in Table III, in terms of their advantages, disadvantages, achievable information transfer rate, and harvested energy under different received power levels and SNR scenarios. Note that a separated information-energy receiver architecture offers the advantage of transmitting the information using an OFDM signal, as the portion of the signal being utilized for power delivery can be separated from the information signal with the help of distinguishing the information and power frequency bands. After the separation of the information and power signals, information processing can be performed as usual. However, this would not be efficient in a practical SWIPT system, as the power required for OFDM signal processing is quite high. Therefore, it is necessary to consider a practical SWIPT scenario where the transmitter and receiver power consumption is also considered. In this direction, the idea of using an integrated information-energy receiver architecture to reduce receiver power consumption is discussed in the next section.

Table III: Transmission designs for a separated receiver architecture.

| Transmission Approach | Advantages | Drawbacks | Information Rate and Energy |
|---|---|---|---|
| OFDM frequency spectrum splitting [43] | • No power splitting or time switching needed. | • Well-designed BPF needed.<br><br>• Linear energy harvester model considered. | 4 bps/Hz and 18 mW for SNR=50 dB. |
| Multisine superimposed on OFDM [45] | • Higher PAPR. | • Saturation of ADC may occur. | 1.8 µA for 1 bit/s/Hz for $SNR = 20$ dB. |
| Rectangular pulse superimposed on CP-OFDM [46] | • No modification in information detection. | • Interference introduced according to the width of the rectangular pulse. | 1.4 V (maximum) at received input power of -15 dBm. |
| Partial usage of information part of OFDM [47] | • No modification in information detection. | • Optimization needed for minimum usage of information part for sufficient harvested energy. | ~38% PCE for 2 bits/subcarrier channel for $SNR = 20$ dB. |
| OFDM-DC signal [48] | • Rectifying components such as diodes are not required.<br><br>• LPF is sufficient to extract the DC. | • Complete redesign of transmitter and receiver. | 2 mW and SER=$10^{-1}$ with 2 bits/channel at $SNR = 10$ dB. |

## 3.2  Integrated Node Architecture

The above-discussed methods are tailored for the separated receiver architecture, in which it is assumed that the receiver has two separate signal streams, namely one for information and one for energy. In the case of an integrated architecture (Fig. 3(b)), the received signal is first passed through the rectifier, and then the resulting baseband signal is divided between two streams, one for energy and one for information, using a voltage divider. As this architecture offers the great advantage of having eliminated the RF mixer, as well as its driving local oscillator, the overall receiver's power consumption reduces by a significant factor. However, new modulation approaches need to be explored for such a receiver architecture.

The simplest modulation technique for an integrated receiver was introduced in 2013 [9]. It adopted a single-tone energy modulation, where information symbols were encoded using different energy levels, also referred to as energy modulation. However, it does not take advantage of multitone signals for increased power performance. To consider these high PAPR multitone signals for integrated node architecture, new techniques for embedding information over these signals and further decoding at the receiver need to be developed.

One way of utilizing high PAPR multitone signals for SWIPT systems is to design the multitone waveform with different PAPR levels for each symbol [51]. As shown in Fig. 11, multiple PAPR levels are produced by varying the number of tones or by varying the tone spacing. A data rate of 0.5 Mbps with $10^{-2}$ bit-error-rate (BER) and a DC output of 3.5 times higher compared to a single tone carrier can be achieved for a 30 dB SNR. However, in [51], the signal BW keeps changing for each symbol, putting additional constraints on the input matching network for large modulation orders. Therefore, PAPR-based information transfer using a fixed BW multitone signal is proposed in [52] where the tone spacing is changed for each transmitted symbol. This approach has been shown to perform better compared to varying the BW multitone signal under low SNR transmission conditions.

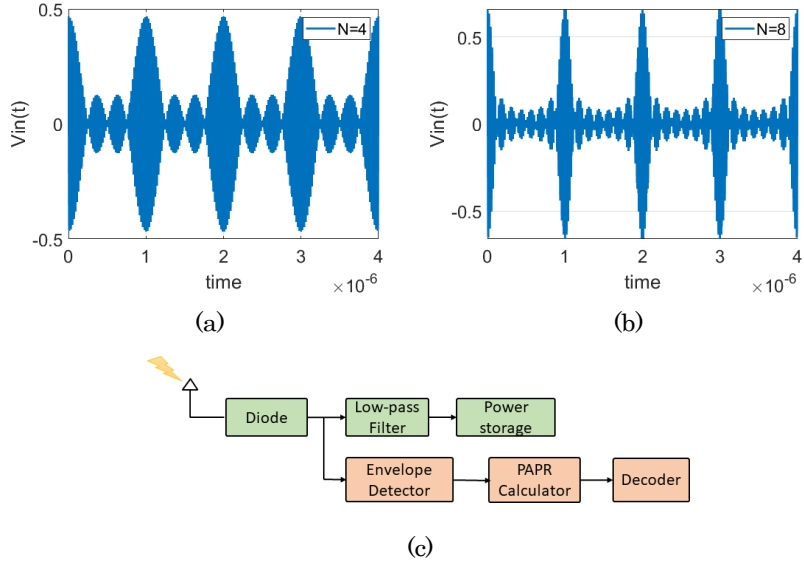

Fig. 11: PAPR-based waveform for (a) $N = 4$, (b) $N = 8$, and its (c) receiver architecture.

These high PAPR-based modulation techniques can be combined with the other single tone based techniques to widen the operating region in order to make the overall SWIPT system more efficient. For example, multitone waveforms outperform the single tone waveform only in the case of a low-power region as the diode enters the saturation region at comparatively higher input power [53]. Therefore, an adaptive receiver structure has been proposed in [54], with a switching option between information paths for single tone and multitone, according to the received input power levels.

Furthermore, in [55], a transmission approach is developed from a transmitter perspective. Here, instead of focusing on RF-DC efficiency at the receiver, the DC-RF efficiency at the transmitter is improved. Hence, a low PAPR signal is designed by superimposing a high power unmodulated continuous wave power signal on the modulated signal, as shown in Fig. 12. In this way, interference between the information and energy signals is minimized by allocating power to the narrowband and information to the wider band. This idea might be feasible in practical scenarios where the power signal generally needs to have much higher power compared to the information signal (i.e., 0 dBm for power and -100 dBm for information).

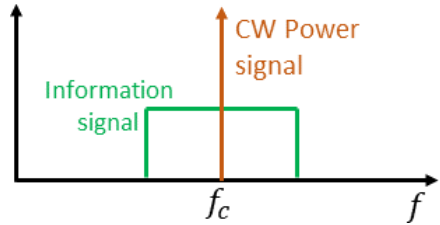

Fig. 12: Frequency spectrum of unmodulated power signal superimposed on modulated information signal.

Generally, most of the research challenges for enhancing PCE in WPT systems revolve around the problem of reducing the ripple present in the output voltage. However, having some amount of ripple may lead to successful decoding of information using the same rectifier hardware, resulting in a trade-off between WIT and WPT. One such technique utilizing these ripples, is to embed information in the amplitude, known as $M$-ary ASK (amplitude-shift-keying) where the amplitude level of the single tone is changed according to the symbol pattern, with each symbol carrying some minimum energy level [56]. Using this modulation method, 0.13 V voltage and a quite low BER of $10^{-4}$ can be achieved for a received input power of $-20$ dBm. Additionally, this technique ensures a minimal continuous power transmission to the receiver as each symbol is carrying some amount of energy. However, this transmission technique uses only a single tone for generating amplitude variations, which degrades the PCE,

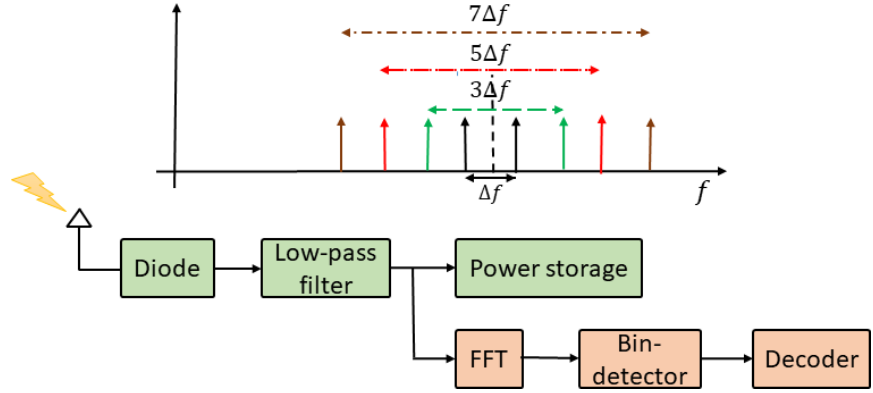

Fig. 13: Multitone FSK waveform and receiver architecture [58].

and does not make use of a high PAPR multitone signal.

A modulation technique that does use a multitone signal is introduced in [57]. Instead of embedding information directly in the amplitudes of the tones, a model is proposed where the information is embedded in the ratios of the amplitudes of the different tones rather than in the amplitudes themselves. Although having information in amplitude ratios offers an advantage of making the system independent of transmission distance while providing a 48% PCE for a received signal power of $-10$ dBm, this technique is only suitable for multitone with smaller numbers of tones as the complete solution of the non-linear equations is needed for proper information decoding, due to the rectifier's non-linearity.

Nonetheless, amplitude variations in the multitone waveform has benefit in terms of data rate (i.e., with the increase of modulation order) for a SWIPT system, but these variations impose a limit on achievable WPT due to the presence of ripples in the output voltage that degrade the WPT performance [56, 57]. An ideal waveform for a SWIPT system would entail minimal variations in the envelope, with the stream of information symbols having a minimum effect upon the WPT. Consequently, a multitone frequency-shift keying (FSK) is proposed in order to reduce the mutual impact of the WPT and WIT on each other [58]. The signal is designed in such a way that different information symbols correspond to different frequency spaced multitones at the transmitter (Fig. 13). Further, decoding at the receiver can be performed in two ways: either by using Fast Fourier Transform (FFT) for identifying the strongest baseband signal, or by evaluating multitone PAPR levels with the changing frequency spacings to lower the power consumption at the receiver [59]. In this way, a 0.55 V output voltage can be attained, which is higher than the amplitude based biased-ASK waveform. However, SER performance degrades to $10^{-2}$ compared to $10^{-4}$ for biased-ASK.

To further increase the data transmission capacity of the integrated receiver architecture, a design using two half-wave rectifiers instead of using one rectifier has been introduced in [60]. Two half-wave rectifiers are utilized with two amplitudes, one positive and one negative, at the output. Different combinations of these two voltage levels, amplitude difference shift keying (ADSK) and amplitude ratio shift keying (ARSK), are realized to increase the available constellation range for the symbols. In short, the allowable modulation order is increased while keeping the same system noise margin, with the help of modification in the receiver architecture. However, the performance is analyzed without taking into consideration the ripples in the DC voltage, which is not a practical scenario while encoding information in the amplitudes.

The various advantages, disadvantages, and performance analysis of the integrated information-energy receiver architecture based transmission approaches are summarized in Table IV. It can be observed that in most of the transmission approaches and for both the receiver architectures, the PCE is enhanced by giving the transmitted signal a higher PAPR by means of a multitone signal. This is similar to the case of the separated information-energy receiver archi-tecture where a high PAPR OFDM signal is used. Indeed, high PAPR signals offer the advantage of increased power transfer performance. However, these high PAPR signals may saturate the non-linear amplifier at the transmitter and subsequently degrade the information transfer performance [61]. Therefore, a complete transmitter-receiver system analysis is required to attain the actual PCE and information transfer performance of these SWIPT transmission schemes.

Table IV: Transmission designs for integrated receiver architecture.

| Transmission Approach | Advantages | Disadvantages | Information and Power |
|---|---|---|---|
| PAPR based [51] | • High PAPR | • Performs well only for high SNR.<br>• Input matching network for a large BW. | 0.5 Mbps and DC of 3.5 times higher compared to the single carrier input signal for 30 dB SNR. |
| PAPR based with fixed signal BW [52] | • Good performance for high SNR as well as low SNR. | • Non-uniform frequency spacings. | 0.5 µA and $BER = 10^{-1}$ for $-10$ dB input power. |
| Biased-ASK [56] | • Each symbol has some minimum energy. | • Single tone is used. | $10^{-4}$ BER and 0.13 V at $-20$ dBm received power, for $SNR = 18$ dB. |
| Amplitude ratio [57] | • Independent of transmission distance. | • Analysis only for the multitone with a smaller number of tones. | 48% PCE for received power of $-10$ dBm. |
| Multitone-FSK [58] | • Lessened envelope variations.<br>• Reduced impact of large WPT on WIT. | • Power consuming FFT needed. | $SER = 10^{-2}$ and 0.55 V for received power of 0 dBm. |
| ADSK, ARSK [60] | • Increased operational range. | • Only DC power is considered. | $10^{-1}$ BER and 0.45 µW at $-10$ dBm received power, for $SNR = 15$ dB. |

The discussed transmission waveforms and the corresponding receiver architectures are still far from practical SWIPT systems. Some of the reasons are: (i) OFDM signal waveforms utilizing a separated receiver architecture consume significant power for signal processing at the receiver, (ii) the effect of WIT over WPT is still significant for the modulation methods used for an integrated information-energy architecture due to the presence of ripples at the output, (iii) moving toward more practical SWIPT systems, the constellation range needs to be increased so that the transmission can support higher modulation orders for increasing information rates, and (iv) an end-to-end performance analysis including the transmitter performance for a particular waveform is required.

## 4 Conclusions and Future Outlook

This article discussed beamforming and optimal waveform design to achieve both information and power transfer over the same signal. Distributing antennas contributes to narrower beams, which increases spectral efficiency, resulting in a reduction of interference with neighboring users, which is advantageous from the communications perspective. Additionally, narrow beam avoids wasting power in non-essential areas, resulting in a power management solution for SWIPT systems. As it is clear that massive MIMO is evolving toward a widely distributed antenna environment to create cell-free systems, it is necessary to rethink antenna deployment and beam steering procedures, particularly when massive MIMO is to be used for communications and power transfer simultaneously.

Further, various modulation techniques tailored for separated and integrated information-energy receiver architectures were discussed. This discussion shows that a trade-off exists between information transfer rate and harvested power for both the considered receiver architectures. These modulation approaches should be further investigated considering more practical and time-varying scenarios. For example, an IoT node may not need to receive information and power continuously. Therefore, it will be necessary to optimize these modulation protocols considering the actual needs of each IoT device at a particular point in time. In fact, it may be possible that data rate is a critical requirement for only a few UEs, whereas charging is urgently needed for other UEs. These respective needs for power and information are likely to vary over time. In conclusion, it may be expected that these studies will result in novel waveform standardization efforts, which is essential for actual deployment of SWIPT in IoT networks.

# References

[1] "Cisco annual internet report (2018–2023) white paper," https://www.cisco.com/c/en/us/solutions/collateral/executive-perspectives/annual-internet-report/white-paper-c11-741490.html, accessed: 2021-06-27.

[2] I. Zhou, I. Makhdoom, N. Shariati, M. A. Raza, R. Keshavarz, J. Lipman, M. Abolhasan, and A. Jamalipour, "Internet of things 2.0: Concepts, applications, and future directions," *IEEE Access*, vol. 9, pp. 70 961–71 012, 2021.

[3] Z. Popovic, "Far-field wireless power delivery and power management for low-power sensors," in *IEEE Wireless Power Transfer (WPT)*, 2013, pp. 1–4.

[4] Y. J. Guo and R. W. Ziolkowsk, *Advanced Antenna Array Engineering for 6G and Beyond Wireless Communications*. John Wiley & Sons, 2021.

[5] Z. Xiang and M. Tao, "Robust beamforming for wireless information and power transmission," *IEEE Wireless Communications Letters*, vol. 1, no. 4, pp. 372–375, 2012.

[6] J. Xu, L. Liu, and R. Zhang, "Multiuser MISO beamforming for simultaneous wireless information and power transfer," *IEEE Transactions on Signal Processing*, vol. 62, no. 18, pp. 4798–4810, 2014.

[7] R. Feng, M. Dai, and H. Wang, "Distributed beamforming in MISO SWIPT system," *IEEE Transactions on Vehicular Technology*, vol. 66, no. 6, pp. 5440–5445, 2017.

[8] R. Zhang and C. K. Ho, "MIMO broadcasting for simultaneous wireless information and power transfer," *IEEE Transactions on Wireless Communications*, vol. 12, no. 5, pp. 1989–2001, 2013.

[9] X. Zhou, R. Zhang, and C. K. Ho, "Wireless information and power transfer: Architecture design and rate-energy tradeoff," *IEEE Transactions on Communications*, vol. 61, no. 11, pp. 4754–4767, November 2013.

[10] A. S. Boaventura and N. B. Carvalho, "Maximizing DC power in energy harvesting circuits using multisine excitation," in *IEEE MTT-S International Microwave Symposium*, June 2011, pp. 1–4.

[11] L. Yang, Y. Zeng, and R. Zhang, "Wireless power transfer with hybrid beamforming: How many RF chains do we need?" *IEEE Transactions on Wireless Communications*, vol. 17, no. 10, pp. 6972–6984, 2018.

[12] Y. J. Guo, M. Ansari, and N. J. G. Fonseca, "Circuit type multiple beamforming networks for antenna arrays in 5G and 6G terrestrial and non-terrestrial networks," *IEEE Journal of Microwaves*, 2021, early access, DOI:10.1109/JMW.2021.3072873.

[13] Y. J. Guo, M. Ansari, W. Z. Richard, and J. G. N. Fonesca, "Quasi-optical multi-beam antenna technologies for B5G and 6G mmWave and THz networks: A review," *IEEE Open J. Antennas Propag.*, 2021, early access, DOI:10.1109/OJAP.2021.3093622.

[14] J. Butler, "Beam-forming matrix simplifies design of electronically scanned antenna," *Electron. Design*, vol. 9, pp. 170–173, 1961.

[15] H. Moody, "The systematic design of the butler matrix," *IEEE Transactions on Antennas and Propagation*, vol. 12, no. 6, pp. 786–788, 1964.

[16] S. Stein, "On cross coupling in multiple-beam antennas," *IRE Transactions on Antennas and Propagation*, vol. 10, no. 5, pp. 548–557, 1962.

[17] W. Kahn and H. Kurss, "The uniqueness of the lossless feed network for a multibeam array," *IRE Transactions on Antennas and Propagation*, vol. 10, no. 1, pp. 100–101, 1962.

[18] C.-H. Tseng, C.-J. Chen, and T.-H. Chu, "A low-cost 60-GHz switched-beam patch antenna array with Butler matrix network," *IEEE Antennas Wireless Propag. Lett.*, vol. 7, pp. 432–435, Jul. 2008.

[19] M. Ansari, H. Zhu, N. Shariati, and Y. Guo, "Compact planar beamforming array with endfire radiating elements for 5G applications," *IEEE Transactions on Antennas and Propagation*, vol. 67, no. 11, pp. 6859–6869, Nov. 2019.

[20] A. Bekasiewicz and S. Koziewl, "Compact 4 x 4 butler matrix with non-standard phase differences for iot applications," *Electronics Letters*, vol. 57, no. 10, pp. 387–389, 2021.

[21] J. Nolen, "Synthesis of multiple beam networks for arbitrary illuminations," Ph.D. Diss., Radio Division, Bendix Corp., Baltimore, MD, USA, Apr. 1965.

[22] P. Li, H. Ren, and B. Arigong, "A symmetric beam-phased array fed by a Nolen matrix using 180° couplers," *IEEE Microw. Wireless Compon. Lett.*, vol. 30, no. 4, pp. 387–390, Apr. 2020.

[23] T. Djerafi, N. J. Fonseca, and K. Wu, "Broadband substrate integrated waveguide 4x4 Nolen matrix based on coupler delay compensation," *IEEE Trans. Microw. Theory Tech.*, vol. 59, no. 7, pp. 1740–1745, May 2011.

[24] W. Lin and R. W. Ziolkowski, "Theoretical analysis of beam-steerable, broadside-radiating huygens dipole antenna arrays and experimental verification of an ultrathin prototype for wirelessly powered iot applications," *IEEE Open Journal of Antennas and Propagation*, vol. 2, pp. 954–967, 2021.

[25] C.-M. Chen, A. P. Guevara, and S. Pollin, "Scaling up distributed massive MIMO: Why and how," in *Asilomar Conference on Signals, Systems, and Computers*, 2017, pp. 271–276.

[26] A. P. Guevara, S. De Bast, and S. Pollin, "Massive MIMO: A measurement-based analysis of MR power distribution," in *IEEE Global Communications Conference (GLOBECOM)*, 2020, pp. 1–6.

[27] T. Lo, "Maximum ratio transmission," *IEEE Transactions on Communications*, vol. 47, no. 10, pp. 1458–1461, 1999.

[28] Q. H. Spencer, A. L. Swindlehurst, and M. Haardt, "Zero-forcing methods for downlink spatial multiplexing in multiuser MIMO channels," *IEEE Transactions on Signal Processing*, vol. 52, no. 2, pp. 461–471, Feb 2004.

[29] E. Björnson, J. Hoydis, and L. Sanguinetti, *Massive MIMO Networks: Spectral, Energy, and Hardware Efficiency*. Foundations and Trends, 2017.

[30] X. Li, E. Björnson, E. G. Larsson, S. Zhou, and J. Wang, "Massive MIMO with multi-cell MMSE processing: exploiting all pilots for interference suppression," *EURASIP Journal on Wireless Communications and Networking*, vol. 2017, no. 1, p. 117, Jun 2017.

[31] S. Claessens, C.-M. Chen, D. Schreurs, and S. Pollin, "Massive MIMO for SWIPT: A measurement-based study of precoding," in *IEEE International Workshop on Signal Processing Advances in Wireless Communications (SPAWC)*, 2018, pp. 1–5.

[32] S. D. Bast, A. P. Guevara, and S. Pollin, "CSI-based positioning in massive MIMO systems using convolutional neural networks," in *IEEE Vehicular Technology Conference (VTC2020-Spring)*, 2020, pp. 1–5.

[33] A. P. Guevara, S. D. Bast, and S. Pollin, "Weave and conquer: A measurement-based analysis of dense antenna deployments," in *IEEE International Conference on Communications (ICC)*, 2021, pp. 1–6.

[34] A. P. Guevara, S. De Bast, and S. Pollin, "MaMIMO user grouping strategies: How much does it matter?" in *Asilomar Conference on Signals, Systems, and Computers*, 2019, pp. 853–857.

[35] A. P. Guevara and S. Pollin, "Densely deployed indoor massive MIMO experiment: From small cells to spectrum sharing to cooperation," *Sensors*, vol. 21, no. 13, 2021.

[36] A. Costanzo, D. Masotti, G. Paolini, and D. Schreurs, "Evolution of SWIPT for the IoT world: Near- and far-field solutions for simultaneous wireless in-

formation and power transfer," *IEEE Microwave Magazine*, vol. 22, no. 12, pp. 48–59, 2021.

[37] C. Lo, Y. Yang, C. Tsai, C. Lee, and C. Yang, "Novel wireless impulsive power transmission to enhance the conversion efficiency for low input power," in *IEEE MTT-S International Microwave Workshop Series on Innovative Wireless Power Transmission: Technologies, Systems, and Applications*, May 2011, pp. 55–58.

[38] N. Shariati, J. R. Scott, D. Schreurs, and K. Ghorbani, "Multitone excitation analysis in RF energy harvesters—considerations and limitations," *IEEE Internet of Things Journal*, vol. 5, no. 4, pp. 2804–2816, 2018.

[39] N. Pan, D. Belo, M. Rajabi, D. Schreurs, N. B. Carvalho, and S. Pollin, "Bandwidth analysis of RF-DC converters under multisine excitation," *IEEE Transactions on Microwave Theory and Techniques*, vol. 66, no. 2, pp. 791–802, Feb 2018.

[40] N. Shariati, W. S. Rowe, J. R. Scott, and K. Ghorbani, "Multi-service highly sensitive rectifier for enhanced RF energy scavenging," *Scientific reports*, vol. 5, p. 9655, 2015.

[41] R. Keshavarz and N. Shariati, "Highly sensitive and compact quad-band ambient RF energy harvester," *IEEE Transactions on Industrial Electronics*, pp. 1–1, 2021.

[42] H. Sakaki, S. Yoshida, K. Nishikawa, and S. Kawasaki, "Analysis of rectifier operation with FSK modulated input signal," in *IEEE Wireless Power Transfer (WPT)*, May 2013, pp. 187–190.

[43] W. Lu, Y. Gong, J. Wu, H. Peng, and J. Hua, "Simultaneous wireless information and power transfer based on joint subcarrier and power allocation in OFDM systems," *IEEE Access*, vol. 5, pp. 2763–2770, 2017.

[44] M. Konstantinos, A. Adamis, and P. Constantinou, "Receiver architectures for OFDMA systems with subband carrier allocation," in *European Wireless Conference*, 2008, pp. 1–7.

[45] B. Clerckx, "Waveform optimization for SWIPT with nonlinear energy harvester modeling," in *International ITG Workshop on Smart Antennas (WSA)*, March 2016, pp. 1–5.

[46] H. Kassab and J. Louveaux, "Simultaneous wireless information and power transfer using rectangular pulse and CP-OFDM," in *IEEE International Conference on Communications (ICC)*, May 2019, pp. 1–6.

[47] R. F. Buckley and R. W. Heath, "System and design for selective OFDM SWIPT transmission," *IEEE Transactions on Green Communications and Networking*, vol. 5, no. 1, pp. 335–347, 2021.

[48] M. N. Khormuji, B. M. Popović, and A. G. Perotti, "Enabling SWIPT via OFDM-DC," in *IEEE Wireless Communications and Networking Conference (WCNC)*, April 2019, pp. 1–6.

[49] A. Rajaram, D. N. K. Jayakody, B. Chen, R. Dinis, and S. Affes, "Modulation-based simultaneous wireless information and power transfer," *IEEE Communications Letters*, vol. 24, no. 1, pp. 136–140, 2020.

[50] E. Bayguzina and B. Clerckx, "Asymmetric modulation design for wireless information and power transfer with nonlinear energy harvesting," *IEEE Transactions on Wireless Communications*, vol. 18, no. 12, pp. 5529–5541, 2019.

[51] D. I. Kim, J. H. Moon, and J. J. Park, "New SWIPT using PAPR: How it works," *IEEE Wireless Communications Letters*, vol. 5, no. 6, pp. 672–675, Dec 2016.

[52] I. Krikidis and C. Psomas, "Tone-index multisine modulation for SWIPT," *IEEE Signal Processing Letters*, vol. 26, no. 8, pp. 1252–1256, Aug 2019.

[53] M. Rajabi, N. Pan, S. Pollin, and D. Schreurs, "Impact of multisine excitation design on rectifier performance," in *2016 46th European Microwave Conference (EuMC)*, 2016, pp. 1151–1154.

[54] J. J. Park, J. H. Moon, K. Lee, and D. I. Kim, "Dual mode SWIPT: Waveform design and transceiver architecture with adaptive mode switching policy," in *IEEE Vehicular Technology Conference (VTC Spring)*, June 2018, pp. 1–5.

[55] K. W. Choi, S. I. Hwang, A. A. Aziz, H. H. Jang, J. S. Kim, D. S. Kang, and D. I. Kim, "Simultaneous wireless information and power transfer (SWIPT) for internet of things: Novel receiver design and experimental validation," *IEEE Internet of Things Journal*, vol. 7, no. 4, pp. 2996–3012, 2020.

[56] S. Claessens, N. Pan, M. Rajabi, D. Schreurs, and S. Pollin, "Enhanced biased ASK modulation performance for SWIPT with AWGN channel and dual-purpose hardware," *IEEE Transactions on Microwave Theory and Techniques*, vol. 66, no. 7, pp. 3478–3486, July 2018.

[57] M. Rajabi, N. Pan, S. Claessens, S. Pollin, and D. Schreurs, "Modulation techniques for simultaneous wireless information and power transfer with an integrated rectifier–receiver," *IEEE Transactions on Microwave Theory and Techniques*, vol. 66, no. 5, pp. 2373–2385, May 2018.

[58] S. Claessens, N. Pan, D. Schreurs, and S. Pollin, "Multitone FSK modulation for SWIPT," *IEEE Transactions on Microwave Theory and Techniques*, vol. 67, no. 5, pp. 1665–1674, May 2019.

[59] T. Ikeuchi and Y. Kawahara, "Peak to average power ratio based signal detection for frequency shift multitone SWIPT system," *IEEE Access*, vol. 9, pp. 4158–4172, 2021.

[60] D. Kim, H. Lee, K. Kim, and J. Lee, "Dual amplitude shift keying with double half-wave rectifier for SWIPT," *IEEE Wireless Communications Letters*, vol. 8, no. 4, pp. 1020–1023, Aug 2019.

[61] J. J. Park, J. H. Moon, H. H. Jang, and D. I. Kim, "Performance analysis of power amplifier nonlinearity on multi-tone SWIPT," *IEEE Wireless Communications Letters*, vol. 10, no. 4, pp. 765–769, 2021.